# Determination of a time-shift in the OPERA set-up using high energy horizontal muons in the LVD and OPERA detectors

**For the LVD Collaboration**
N.Yu. **Agafonova** [a], P. **Antonioli** [b], V.V. **Ashikhmin** [a], G. **Bari** [b], E. **Bressan** [b], L. **Evans** [c], M. **Garbini** [b,f], P. **Giusti** [b], A.S. **Malguin** [a], R. **Persiani** [b,f], V.G. **Ryasny** [a], O.G. **Ryazhskaya** [a], G. **Sartorelli** [b,f], E. **Scapparone** [b], M. **Selvi** [b], I.R. **Shakirianova** [a], L. **Votano** [d,e], H. **Wenninger** [c], V.F. **Yakushev** [a], A. **Zichichi** [b,c].
*a) INR-RAS, Moscow, Russia*
*b) Bologna INFN, Bologna, Italy*
*c) CERN, Geneva, Switzerland*
*d) INFN-LNF Frascati, Italy*
*e) INFN-LNGS, Assergi, Italy*
*f) Bologna University, Bologna, Italy*

**For the OPERA Collaboration**
N. **Agafonova** [a], A. **Alexandrov** [b], A. **Bertolin** [c], R. **Brugnera** [d,c], B. **Büttner** [e], V. **Chiarella** [f], A. **Chukanov** [g], N. **D'Ambrosio** [h], G. **De Lellis** [i,b], A. **Di Crescenzo** [i,b], D. **Di Ferdinando** [l], N. **Di Marco** [h], S. **Dmitrievsky** [g], M. **Dracos** [m], S. **Dusini** [c], J. **Ebert** [e], A. **Ereditato** [n], T. **Ferber** [e], R.A. **Fini** [o], A. **Garfagnini** [d,c], G. **Giacomelli** [p,l], C. **Göllnitz** [e], Y. **Gornushkin** [g], F. **Grianti** [q], C. **Gustavino** [v], C. **Hagner** [e], M. **Hierholzer** [e], A. **Hollnagel** [e], K. **Jakovcic** [r], C. **Jollet-Meregaglia** [m], B. **Klicek** [r], U. **Kose** [c], J. **Lenkeit** [e], A. **Ljubicic** [r], A. **Longhin** [f], A. **Malgin** [a], G. **Mandrioli** [l], V. **Matveev** [a], N. **Mauri** [f], E. **Medinaceli** [d,c], A. **Meregaglia** [m], M.T. **Muciaccia** [t,o], D. **Naumov** [g], A. **Olshevsky** [g], A. **Paoloni** [f], A. **Pastore** [o], L. **Patrizii** [l], M. **Pozzato** [p,l], F. **Pupilli** [h], G. **Rosa** [u,v], I. **Rostovtseva** [z], A. **Russo** [b], O. **Ryazhskaya** [a], A. **Schembri** [h], I. **Shakirianova** [a], A. **Sheshukov** [b], S. **Simone** [t,o], M. **Sioli** [p,l], C. **Sirignano** [d,c], G. **Sirri** [l], M. **Spinetti** [f], L. **Stanco** [c], M. **Stipcevic** [r], M. **Tenti** [p,l], F. **Terranova** [s,f], V. **Tioukov** [b], L. **Votano** [f,h], B. **Wonsak** [e], V. **Yakushev** [a], Y. **Zaitsev** [z], S. **Zemskova** [g].
*a) INR-RAS, Moscow, Russia*
*b) Napoli INFN, Napoli, Italy*
*c) Padova INFN, Padova, Italy*
*d) Padova University, Padova, Italy*
*e) Hamburg University, Hamburg, Germany*
*f) LNF, Frascati, Italy*
*g) JINR, Dubna, Russia*
*h) LNGS, Assergi, Italy*
*i) Napoli University, Napoli, Italy*
*l) Bologna INFN, Bologna, Italy*
*m) IPHC, Strasbourg, France*
*n) LHEP, Bern, Switzerland*
*o) Bari INFN, Bari, Italy*
*p) Bologna University, Bologna, Italy*
*q) Urbino University, Urbino, Italy*
*r) IRB, Zagreb, Croatia*
*s) Milano INFN, Milano, Italy*
*t) Bari University, Bari, Italy*
*u) Roma University, Roma, Italy*
*v) Roma INFN, Roma, Italy*
*z) ITEP, Moscow, Russia*

# Determination of a time-shift in the OPERA set-up using high energy horizontal muons in the LVD and OPERA detectors

**Abstract**

The purpose of this work is to report the measurement of a time-shift in the OPERA set-up in a totally independent way from Time Of Flight (TOF) measurements of CNGS neutrino events and without the need of knowing the distance between the two laboratories, CERN and LNGS, where the neutrinos are produced and detected, respectively. The LVD and OPERA experiments are both installed in the same laboratory: LNGS. The relative position of the two detectors, separated by an average distance of ~ 160 m, allows the use of very high energy horizontal muons to cross-calibrate the timing systems of the two detectors, using a time of flight (TOF) technique, which, as stated above, is totally independent from TOF of CNGS neutrino events. Indeed, the OPERA-LVD direction lies along the so-called "Teramo anomaly", a region in the Gran Sasso massif where LVD has established, many years ago, the existence of an anomaly in the mountain structure, which exhibits a low *m. w. e.* thickness for horizontal directions. The "abundant" high-energy horizontal muons (nearly 100 per year) going through LVD and OPERA exist because of this anomaly in the mountain orography.

The total live time of the data in coincidence between the two experiments correspond to 1200 days from mid 2007 until March 2012.

The time coincidence study of LVD and OPERA detectors is based on 306 cosmic horizontal muon events and shows the existence of a negative time shift in the OPERA set-up of the order of $\Delta t_{AB} = - (73 \pm 9)$ ns when two calendar periods, A and B, are compared. The first, A, goes from August 2007 to August 2008 plus the period from January 2012 to March 2012; the second period, B, goes from August 2008 to December 2011.

This result shows a systematic effect in the OPERA timing system present from August 2008 until December 2011. The size of the effect, in terms of the cosmic horizontal muons TOF, is comparable with the neutrino velocity excess recently measured by OPERA.

It is probably interesting not to forget that with the MRPC technology developed by the ALICE Bologna group the TOF world record accuracy of 20 ps was reached. This technology can be implemented at LNGS for a high precision determination of TOF with the CNGS neutrino beams.

If new experiments are needed for the study of neutrino velocities they must be able to detect effects an order of magnitude smaller than the value of the OPERA systematic effect.

# 1. Introduction

This work is based on the existence at Gran Sasso of very high energy horizontal muons which go through the OPERA and LVD detectors, both installed in the same laboratory, LNGS, and separated by an average distance of about 160 m. The high-energy cosmic horizontal muons allow cross-calibrating the timing systems of the two detectors in a totally independent way from TOF measurements of CNGS neutrino events. Furthermore, our new method has the basic advantage of being totally independent from the knowledge of the long distance of ~730 km between the two Labs: CERN, where the neutrinos are produced and LNGS, where they are detected.

The purpose of the present work is to establish if a time-shift of the OPERA detector, suggested by measurements recently conducted by OPERA on the timing system of the experiment, was also present in the past years, and could be the reason for the reduction of 60 ns in the reported neutrino time of flight.

The original purpose of the search for time coincident events between LVD and OPERA experiments, separated by an average distance of about 160 m, was to perform an unprecedented analysis of very large cosmic ray showers looking at their penetrating TeV component. The physics case follows the consideration that TeV muons separated by hundreds of meters are produced in high $p_T$ ($p_T > 3$ GeV/c) interactions up in the atmosphere.

In a time-window of 1 μs, and excluding events in time with the CNGS beam spill, we found 306 events due to cosmic ray muons. This sample has a time-difference ($\delta t = t_{LVD} - t_{OPERA}$) distribution peaked at 616 ns with an RMS of ~ 74 ns. The central value of the distribution has the following interpretation: the events in coincidence detected up to now are not multiple muons (one per each detector), but single muon events entering horizontally from the OPERA side and going through the LVD detector after a delay of 616 ns (the delay is a sum of Time of Flight plus a stable systematic error). Indeed, the OPERA-LVD direction lies along the so-called "Teramo anomaly", where LVD has established many years ago [1] the existence of an anomaly in the mountain profile which exhibits a low *m. w. e.* (meter, water, equivalent) depth for horizontal directions. Visual inspection using the event displays of both experiments confirms this "anomaly", which allows high-energy horizontal cosmic muons to go through the OPERA and LVD set-ups. These high-energy muons are used to cross-calibrate the two detectors timing systems, through a time of flight (TOF) technique. It is probably interesting to point out that using with the MRPC technology developed by the ALICE Bologna group we obtained the TOF world record of 20 ps [2]. If needed, in the near future, this technology could be implemented at LNGS for a high precision determination of TOF with the CNGS neutrino beam [3].

In chapters 2 and 3, a brief description of the LVD and of the OPERA detectors is reported; chapter 4 is dedicated to the data selection from the two set-ups. In chapter 5 we report the basic principle of the data analysis, whose data were collected from mid 2007 until 2012. The total number of horizontal cosmic muons detected by LVD and OPERA in coincidence is 306 and corresponds to a live time of about 1200 days.

Chapter 6 is dedicated to the evidence for the time shift in the OPERA set-up. Looking at the calendar time evolution of the time difference δt in separate periods of data taking, the average values change. The observed variations are larger than the statistical uncertainty estimated in each period. The results of this joint analysis is the first quantitative measurement of the stability of the relative timing between the two detectors and it establishes the existence of systematic effects in the OPERA detector independently of the TOF measurement of CNGS neutrino events. The stability in time of LVD shows that the OPERA detector has a negative time shift in the calendar period (B) from August 2008 to December 2011 of the order of $\Delta t_{AB} = (-73 \pm 9)$ ns compared with the previous period A (calendar time from August 2007 to August 2008) and the following one (from January 2012 to March 2012) taken together. This effect explains the previous OPERA findings of a neutrino time of flight shorter by 60 ns over the speed of light.

The Conclusions are given in chapter 7. If new experiments will be needed for the study of neutrino velocities they must be able to detect effects an order of magnitude smaller than the value of the detected OPERA systematic effect [3].

## 2. The LVD detector

The Large Volume Detector (LVD) [4] (Figure 1) is located in Hall A of the INFN underground Gran Sasso National Laboratory at an average depth of 3600 *m. w. e.* The LVD main purpose is to detect and study neutrino bursts from galactic gravitational stellar core collapses. The experiment started taking data in June 1992 and has continued without interruptions until now.

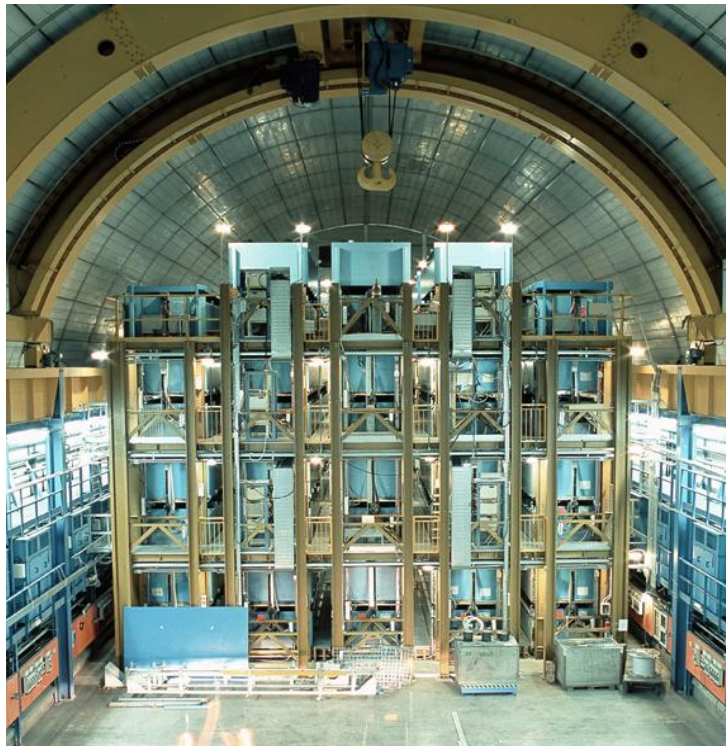

*Fig. 1. Photograph of the LVD detector in the Hall A of the underground Gran Sasso Labs.*

The detector, schematically shown in Figure 2, consists of an array of 840 liquid scintillator counters, 1.5 $m^3$ each, arranged in a compact and modular geometry: 8 counters are assembled in a module called "portatank"; 35 portatanks (5 columns × 7 levels) form a "tower"; the whole detector consists of three identical towers that have independent power supply, trigger and data acquisition systems. The external dimensions of the active volume are $13 \times 23 \times 10$ $m^3$. The liquid scintillator (density $\rho = 0.8$ $g/cm^3$) is $C_nH_{2n}$ with $<n> = 9.6$ doped with 1 g/l of PPO (scintillation activator) and 0.03 g/l of POPOP (wavelength shifter). The total active scintillator mass is M = 1000 t. Each LVD counter is viewed from the top by three 15 cm diameter photomultipliers (FEU49 or FEU125).

The main reaction that is detected by LVD is the inverse beta decay (anti-$\nu_e$ p, n$e^+$), which gives two signals: a prompt one due to the $e^+$ followed by the signal from the neutron capture reaction (np,d$\gamma$) with mean capture time of about 185 μs and $E_\gamma$ = 2.2 MeV.

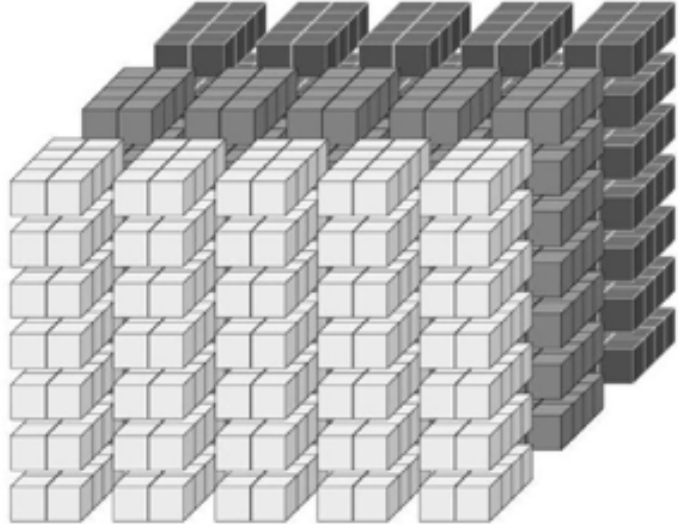

*Fig. 2. Schematic view of the LVD apparatus.*

The trigger logic is optimized for the detection of both products of the inverse beta decay reaction and is based on the three-fold coincidence of a counter PMTs. Each PMT is discriminated at two different levels, $E_{High} \approx 4$ MeV and $E_{Low} \approx 0.5$ MeV, resulting in two possible levels of coincidence between the three PMTs: High Energy Threshold (HET) and Low Energy Threshold (LET). A HET coincidence signal in any counter represents the trigger condition for surrounding counters. Once the trigger counter has been identified the charge of the 3 PMTs summed signals and the time of their coincidence are stored in a memory buffer. For all LET coincidences occurring in anyone of the 8 counters in the same module of the trigger counter within 1 ms from the trigger, the charge of the three PMTs summed signals and the coincidence time are also recorded. Starting 1 ms after the occurrence of a trigger, the read out of the memory buffers, one per module, containing the charge and time information of both HET and LET signals, is performed independently on the three towers without introducing any dead time.

The modularity of the apparatus allows for calibration, maintenance and repair interventions without major negative interference with data taking and detector sensitivity. Figure 3 shows the duty cycle and the trigger active mass of LVD from June 1992 to March 2011. From 2001 the experiment has been in very stable conditions with duty cycle > 99% and slightly increasing active mass. The minimum trigger mass of 300 t, corresponding to less than one "tower", at which LVD can monitor the whole Galaxy for gravitational core collapses is also shown (blue).

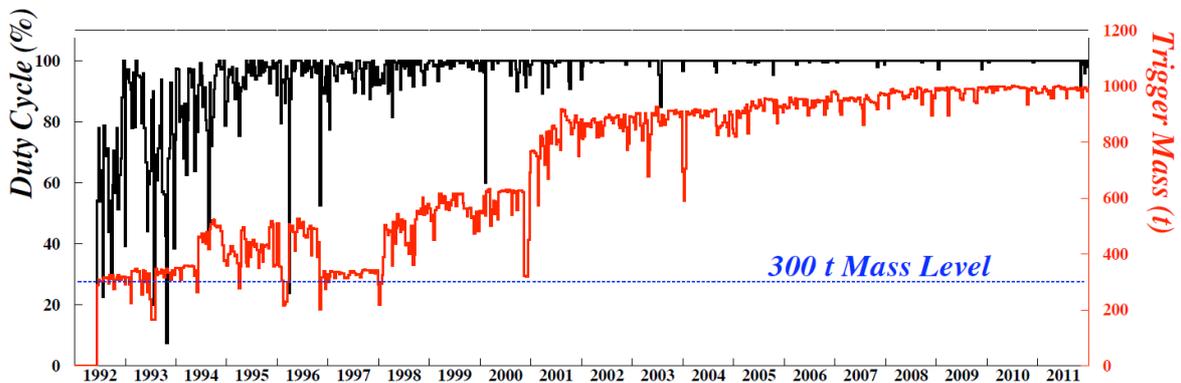

*Fig. 3. LVD duty cycle and active mass in the period June 1992 – March 2011.*

## 3. The OPERA detector

OPERA is a hybrid experiment with electronic detectors and nuclear emulsions located in Hall C of the underground Gran Sasso Laboratory [5]. The main physics goal of the experiment is to observe neutrino flavor oscillations through the appearance of $v_\tau$ neutrinos in the $v_\mu$ CNGS beam. The detector design was optimized to identify the $\tau$ lepton via the topological observation of its decay: this requires a target mass of more than a kton to maximize the neutrino interaction probability and a micrometric resolution to detect the $\tau$ decay. To accomplish these requirements the detector concept is based on the Emulsion Cloud Chamber (ECC) technique combined with real-time detection techniques (electronic detectors). The OPERA detector basic unit, the brick, is a sandwich of 56 lead plates 1 mm thick, interleaved with emulsion films (2 emulsion layers, 44 μm thick, poured on a 205 μm plastic base).

The OPERA apparatus (Figure 4) is divided in two identical supermodules (SM). Each SM has a target part, composed by 31 vertical walls, perpendicular to the beam direction, containing the ECC bricks. The walls are interleaved with planes of plastic scintillator (TT). The two targets contain a total of ~150000 bricks for a mass of ~1.3 ktons. Each target is followed by a magnetic spectrometer consisting of an iron magnet equipped with plastic Resistive Plate Counters (RPC) and vertical drift tubes (Precision Trackers, PT). Finally, two glass RPC planes mounted in front of the first target (VETO) allow rejecting charged particles originating from outside the target fiducial region, because of neutrino interactions occurring in the surrounding rock material.

The electronic detector has the following tasks: i) to provide a neutrino interaction trigger and the event timing; ii) to locate the bricks where interactions occur; iii) to perform the tracking/ identification of muons and measure their charge and momentum; iv) to provide some rough calorimetric measurements of the hadronic energy of the events.

The detector is equipped with a robotized system (the Brick Manipulator System, BMS) that allows removing the bricks with neutrino interactions from the target. A detailed description of the detector and of the data acquisition system can be found in [5].

Events induced by CNGS neutrinos are selected on a delayed time coincidence between the proton extraction from SPS and the neutrino interaction in OPERA. The synchronization is based on a GPS system, with a period of synchronization ~ 100 ns. In Figure 5 the time distribution of CNGS neutrino events is shown [6]. When a CNGS neutrino interacts in OPERA, the event is recorded and reconstructed by the electronic detectors. If there is a muon in the event, its trajectory is traced back through the scintillator planes down to the brick where the track originates. When no muons are observed the scintillator signals produced by electromagnetic or hadronic showers are used to predict the brick containing the neutrino interaction primary vertex. The selected brick is then extracted from the target by the BMS. The overall procedure minimizes the target mass losses, allowing at the same time a semi-online analysis.

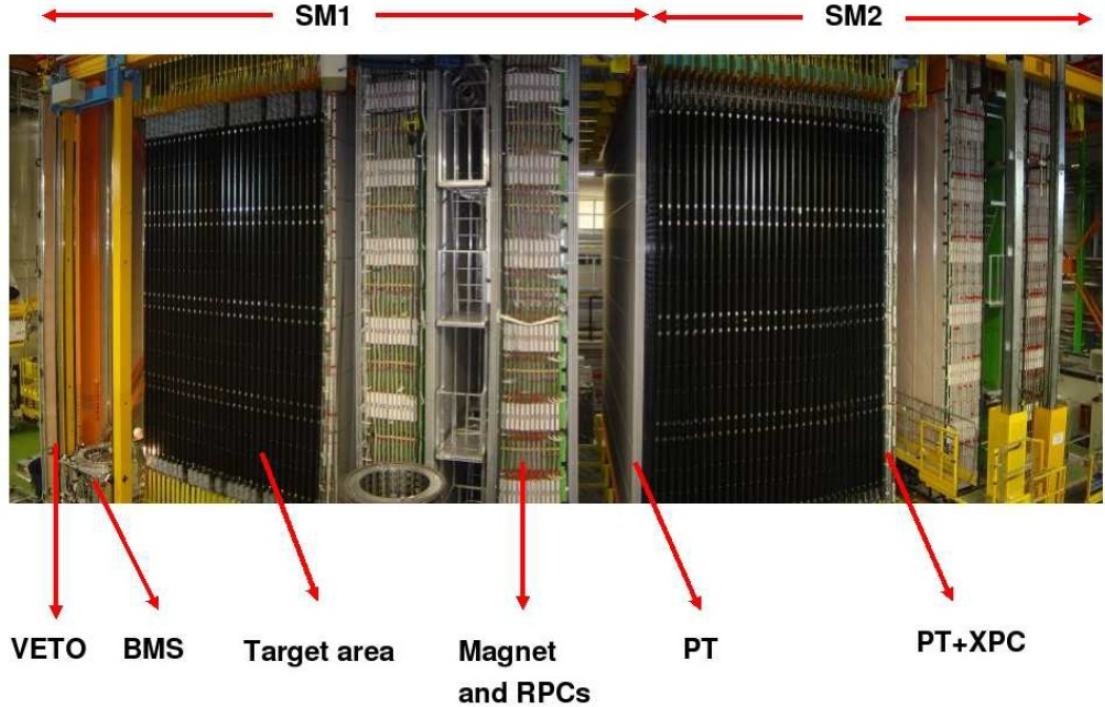

*Fig. 4. Fish-eye side view of the OPERA detector. The upper red horizontal lines indicate the position of the two identical supermodules (SM1 and SM2). The "target area" is made of walls filled with ECC bricks interleaved with planes of plastic scintillators (TT): the black covers are the end-caps of the TT. Arrows also show the position of the VETO planes, the drift tubes (PT) surrounded by the XPC, the magnets and the RPC installed between the magnet iron slabs. The Brick Manipulator System (BMS) is also visible. More details are given in [5].*

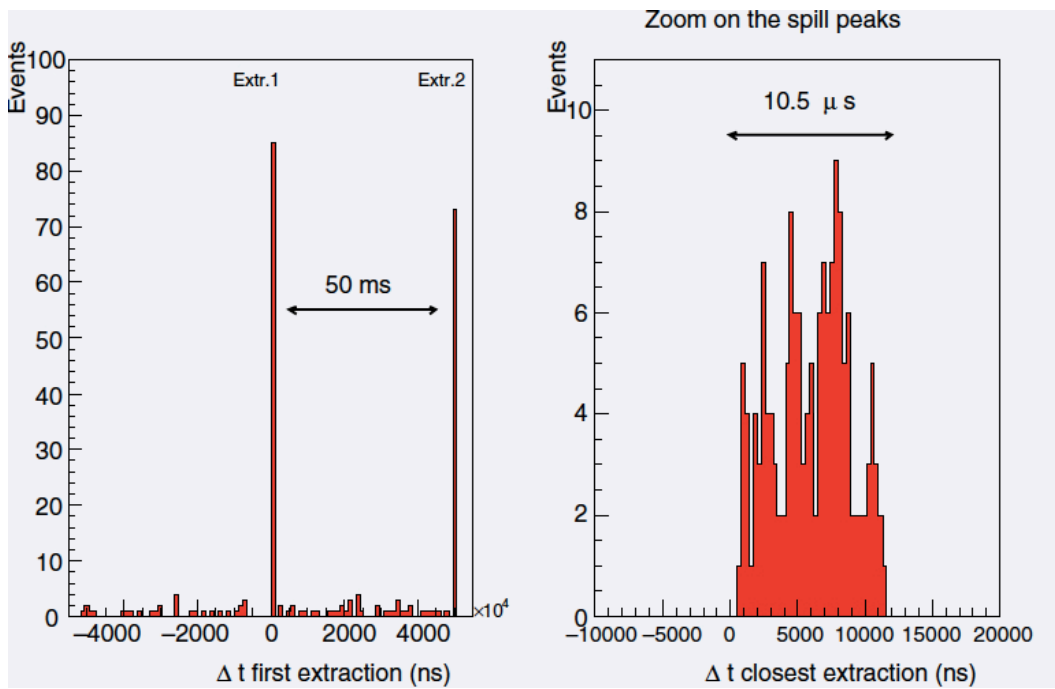

*Fig. 5. Time distribution of events collected in the neutrino run. The event time difference with respect to the closest extraction is shown in the right histogram.*

## 4. Data selection

Muon events have been selected in both LVD and OPERA experiments.

### 4.a. For the LVD detector

In LVD an event is defined to be a muon if:
- there are at least two hit counters, belonging to the sample of GOOD counters (i.e. not high frequency, nor with known problems; this information is defined for each run and saved in the LVD database),
- each of the two counters has an energy deposit of at least 10 MeV.

The LVD detector is equipped with 3 slave clocks (one per tower), each one connected to the external master clock through optical fibers, whose lengths ($\Delta t_{fiber}$) were measured via a two-way fiber measurement (in 2006) and a transportable Cs clock (2007). When a trigger occurs (one counter with an energy release larger than the HET) a signal is sent to the slave clock to read the current time ($t_{clock}$) and save it in the data stream. The time delay between the outputs of the PMTs and the slave clock ($\Delta t_{cable}$) (due to the length of the cables and to the electronic modules needed to form the trigger) was measured in 2009 with a digital scope.

The UTC time of the event in LVD is thus defined as:

$$t_{LVD} = t_{clock} + \Delta t_{fiber}(i) + \Delta t_{cable}$$ where i=1, 2, 3 indicate the tower number and

- $\Delta t_{fiber}(1)$ = 42516 ns (Tower 1)
- $\Delta t_{fiber}(2)$ = 42464 ns (Tower 2)
- $\Delta t_{fiber}(3)$ = 42441 ns (Tower 3)
- $\Delta t_{cable}$ = –350 ns

The reason why $\Delta t_{cable}$ is negative is because in LVD the physical event happens before the stop comes to the slave clocks while in OPERA the timestamp is built at the sensor level.

The contribution to the delay from the muon passage in the counter up to the PMT exit is not taken into account in the present analysis since it is a LVD-constant and it does not alter the time difference between LVD and OPERA as a function of the calendar time. It is nevertheless currently under measurement in view of the next CNGS beam run.

The LVD time accuracy is due mainly to the clock granularity (100 ns) and to the fluctuations in the response of the various PMTs. An analysis performed using the time difference of CNGS neutrino-induced muons crossing two towers allowed to estimate the latter contribution to be of the order of ~35 ÷ 40 ns. The global time accuracy of each event detected by LVD is thus about 50 ns.

We proved the stability of the time performances of LVD studying, with the same technique, the relative time difference between the 3 LVD clocks, along the whole period of time considered in this analysis.

## 4.b. For the OPERA detector

Only events with the earliest hit from the TT or RPC systems were considered, with a minimum number of 20 fired hits.

In order to reconstruct the event time ($t_{OPERA}$) proper corrections are applied depending on the interested subdetector (TT or RPC):

$$t_{OPERA} = t_{clock} + \Delta t_{fiber} + \Delta t_{elec} + \Delta t_{FPGA} + \Delta t_{TT}$$

The different terms contributing to $t_{OPERA}$ are discussed in [7]. Here we briefly recall that:
- $t_{clock}$ is the number frozen by the GPS ESAT clock before the transmission underground
- $\Delta t_{fiber}$ is the delay in propagating the time signal along the 8 km optical fiber from the external GPS clock to the OPERA Master Clock. The $\Delta t_{fiber}$ delay was measured in 2006 and 2007 with a double-path technique and a transportable Cs clock device, respectively, giving comparable results. Further measurements were performed in December 2011 and February 2012 [8].
- $\Delta t_{elec}$ is the delay in transmitting the time base of the OPERA Master Clock to the TT front-end cards
- $\Delta t_{FPGA}$ is internal delay of the FPGA (Field Programmable Gate Array), processing the master clock signal and performing the front-end card time-stamp
- $\Delta t_{TT}$ is the delays in producing the Target Tracker signal including the scintillator response, the propagation of the signals in the WLS (Wave Length Shifter) fibers, the transit time of the photomultiplier and the time response of the OPERA analogue front-end readout chip. The $\Delta t_{TT}$ delay correction has a component (9.4 ns) due to the signal propagation along the scintillator strips. This correction was originally evaluated for CNGS muons which randomly impact the strips in the transverse coordinate. In this work we used a different correction with respect to [7] (53.1 ns instead of 59.6 ns) which takes into account the different arrival direction of Teramo muons with respect to CNGS neutrino induced muons.

It was determined that:
- $\Delta t_{fiber} + \Delta t_{elec} = 40996$ ns + 4262.9 ns [7]
- $\Delta t_{FPGA} + \Delta t_{TT} = 24.5$ ns – 53.1 ns
- If the earliest hit is from RPC → $t_{OPERA} = t_{OPERA} - 148.8$ ns [9]

The final correction to be applied to the OPERA time-stamps is therefore 45230.3 ns for TT and 45081.0 ns for RPC.

## 4.c. Data selection for LVD and OPERA

The UTC time of each muon event is evaluated and exchanged between the two experiments. No other information about the event is transmitted.

We analyzed data collected from summer 2007 up to March 2012. In Table 1 the number of days in common between the two detectors is reported for each running period. Data corresponding to 1207 days were analyzed. In the same Table the event rate, mainly from cosmic muons, for each detector is also reported. In steady conditions the event rate scales with the detector acceptances.

| Period | OPERA | | LVD | | Days in common | Rate OPERA (Hz) | Rate LVD (Hz) | Expected in ±1 ms | Observed in ±1 ms | Expected Teramo m | Observed Teramo m |
|---|---|---|---|---|---|---|---|---|---|---|---|
| 2007 | 28/08/2007 | 31/12/2007 | 27/08/2007 | 31/12/2007 | 58.2 | 0.184 | 0.095 | 177.1 | 162 | 15.7 | 21 |
| 2008 | 01/01/2008 | 05/12/2008 | 01/01/2008 | 07/12/2008 | 263.7 | 0.073 | 0.095 | 314.2 | 323 | 71.2 | 64 |
| 2009 | 01/06/2009 | 23/11/2009 | 31/05/2009 | 01/12/2009 | 171.1 | 0.124 | 0.098 | 359.0 | 351 | 46.2 | 49 |
| 2010 | 31/12/2009 | 31/12/2010 | 01/01/2010 | 01/01/2011 | 326.5 | 0.063 | 0.097 | 346.3 | 369 | 88.2 | 63 |
| 2011 | 31/12/2010 | 07/12/2011 | 01/01/2011 | 01/01/2012 | 336.9 | 0.063 | 0.098 | 360.6 | 395 | 91.0 | 109 |
| 2012 | 09/01/2012 | 02/03/2012 | 01/01/2012 | 03/03/2012 | 50.8 | 0.051 | 0.094 | 41.9 | 37 | 11.1 | 9 |
| | | | | | 1207.2 | | | 1599.1 | 1637 | 323.3 | 315 |

*Table 1. Total statistics analyzed so far for the OPERA-LVD coincidence study. For each year (first column) the corresponding periods for OPERA and LVD, the number of days in common and the event rate (in Hertz) are reported. Note that the LVD rate is in general higher than OPERA rate due to the different detector acceptances.*

In Figure 6 we show the muon rate of the two experiments, taken from the datasets, which have been exchanged. The muon rate in LVD is very stable with an average value (when the apparatus is fully operational) of ~0.1 Hz; in the data sample provided by OPERA there are some spikes, especially during the first three years, then the rate becomes more stable, with an average value of about 0.06 Hz.

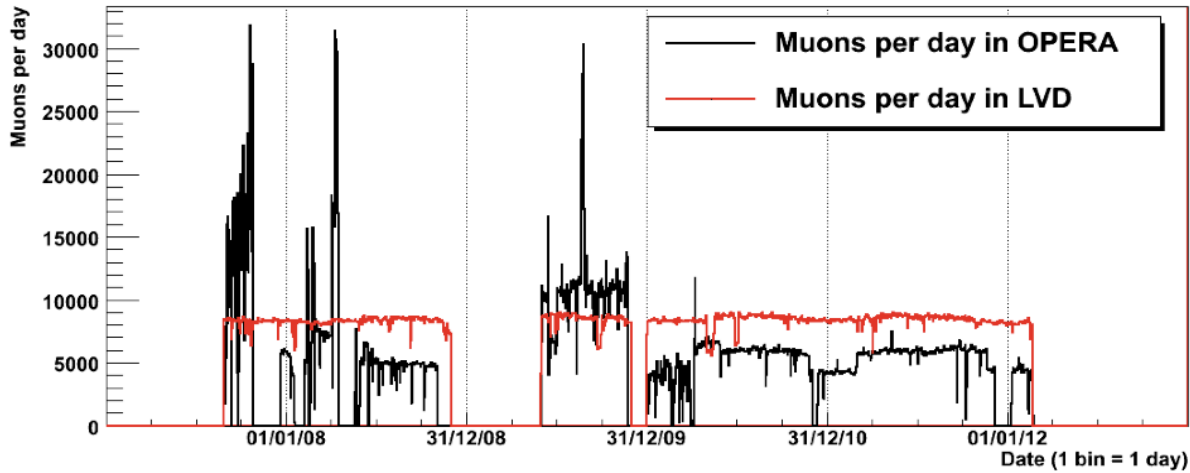

*Fig. 6. Daily "muon" rate of the LVD (red) and OPERA (black) experiments, taken from the datasets, which have been exchanged.*

## 5. Data analysis

Starting from the datasets shown in Figure 6 we look for events in time coincidence between the two experiments. In total 1637 coincident events were found in a time window of 1 ms. The number of events in each year is reported in Table 1 together with the expected number of coincident events computed according to the formula $N_{random} = \mu_{LVD} \mu_{OPERA} \Delta\tau \Delta T$, where $\mu_{LVD}$ and $\mu_{OPERA}$ are the events rates, $\Delta\tau$ is the time window and $\Delta T$ the number of days in common.

In Figure 7 we show the distribution of the time difference

$$\delta t \equiv t_{LVD} - t_{OPERA}$$

for the whole data sample in the ±1 ms time window (left plot) and a zoom in the CNGS spill width (10.5 μs) time window (right plot). Two main components are present: 147 CNGS neutrino events from the same spill (Figure 8, left plot) and 315 events outside the beam spill (Figure 8, right plot).

The 147 CNGS events are pairs of neutrino events coming from the same spill and are distributed according to an approximate triangular function centered on $\delta t = 0$.

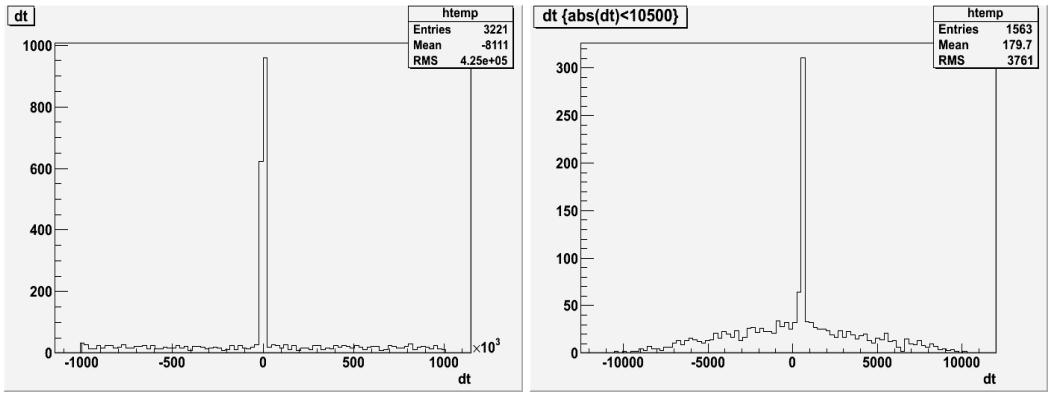

Fig. 7. Distribution of the time difference $\delta t \equiv t_{LVD} - t_{OPERA}$ (left) and a zoom on ±10.5 μs (right). Units are microseconds.

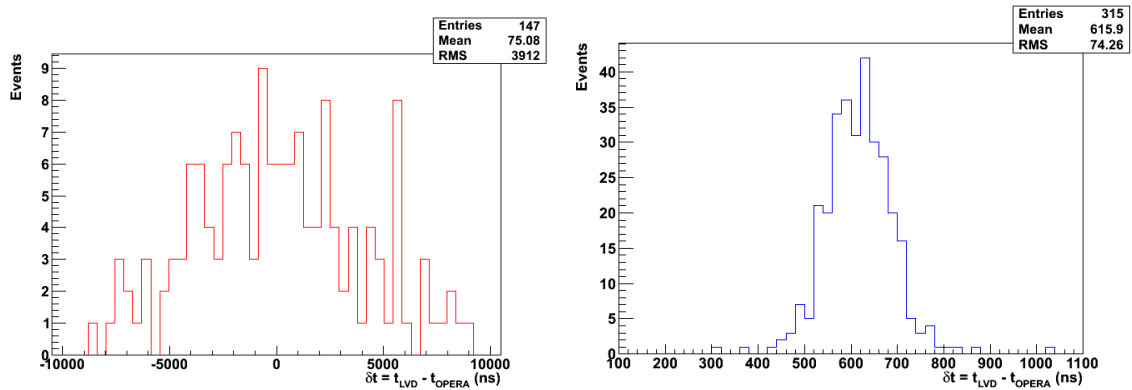

Fig. 8. Distribution of the time difference $\delta t$ for CNGS muon events coming from the same neutrino spill (left), and time difference $\delta t$ for "cosmics" (right).

We now consider the 315 events outside the CNGS beam spill time ("cosmic" events). The results obtained with a large coincident window of ±10 μs are shown as a function of the calendar time in Figure 9. A highly-populated region appears for the whole dataset at time difference $\delta t$ between 0 and 1 μs. The zoom of this region is shown in Figure 10.

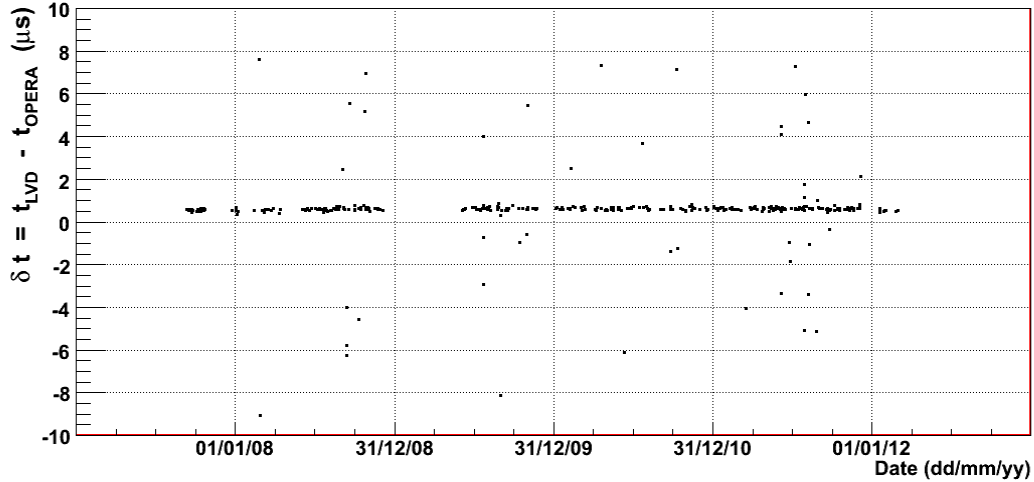

*Fig. 9. Time difference between the UTC time of the LVD and OPERA experiments for "cosmic" events as a function of the calendar date.*

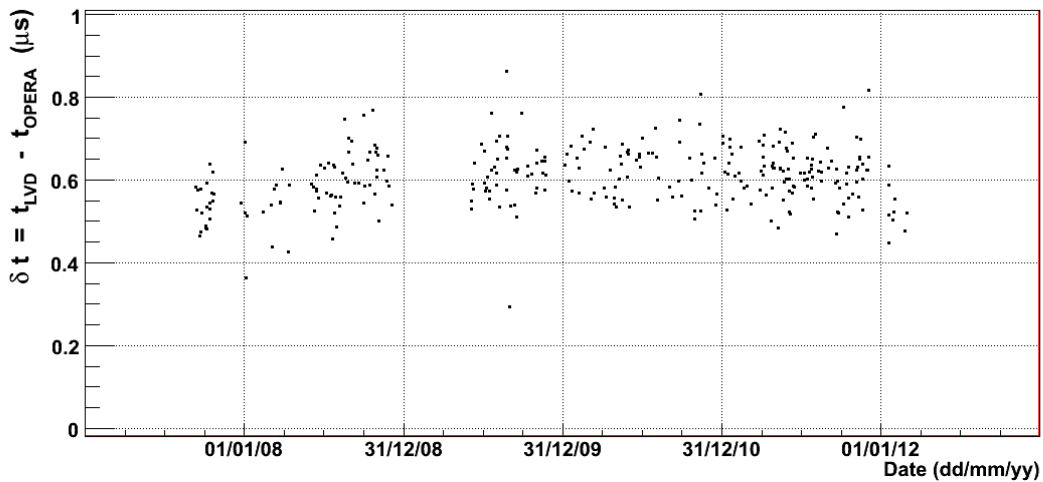

*Fig. 10. Time difference between the UTC time of the LVD and OPERA experiments for "cosmic" events as a function of the calendar date: zoom for δt in [0; 1] μs.*

The average value of δt over the whole sample is 616 ns, as shown in Figure 8 (right). This value is not compatible with multiple muon events hitting both LVD and OPERA from above. On the other hand it is compatible with a horizontal muon, directed along the highway tunnel, hitting OPERA first and then LVD, which are about 160 m far away. A sketch of the LNGS map with the position of LVD and OPERA experiments is shown in Figure 11. Indeed, the OPERA-LVD direction lies along the anomaly in the mountain profile observed in 1997 [1] when searching for neutrino events from the center of the Galaxy. The anomaly is due to the non-uniform depth of the rock structure in the horizontal direction towards the city of Teramo, thus called "Teramo anomaly". This is due to a large decrease in *m. w. e.* of the mountain rock structure, as indicated in Figure 12 by the red circle.

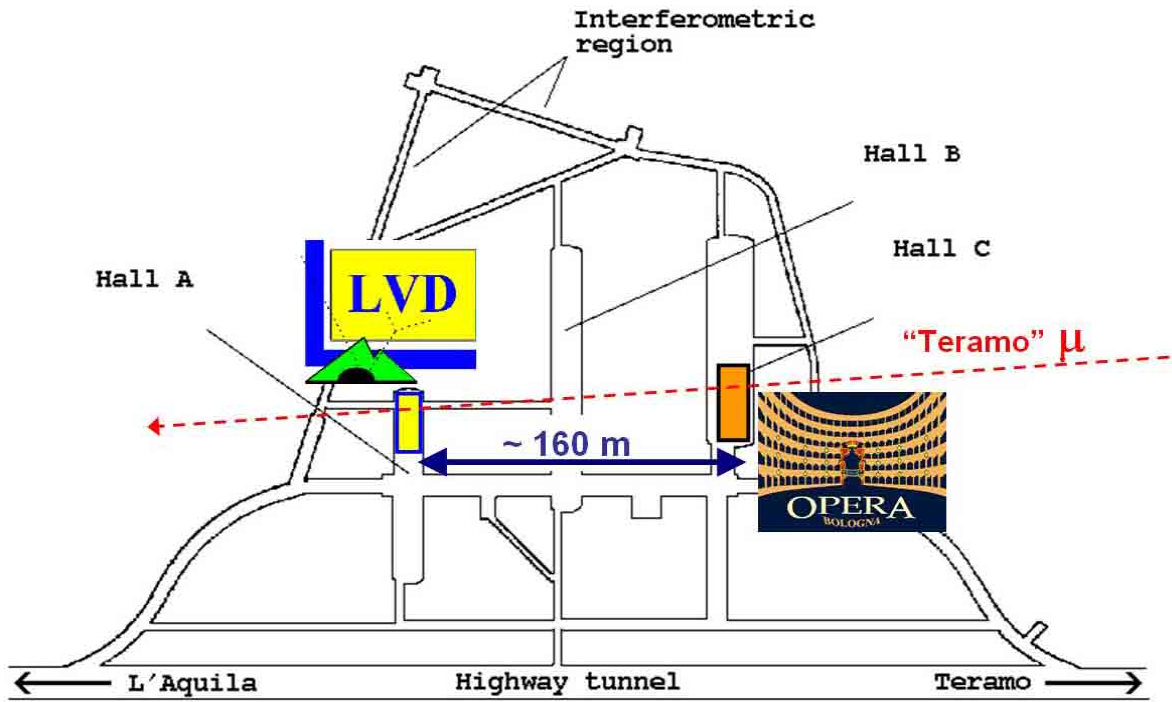

*Fig. 11. Sketch of the LNGS map with the position of the LVD and OPERA experiments.*

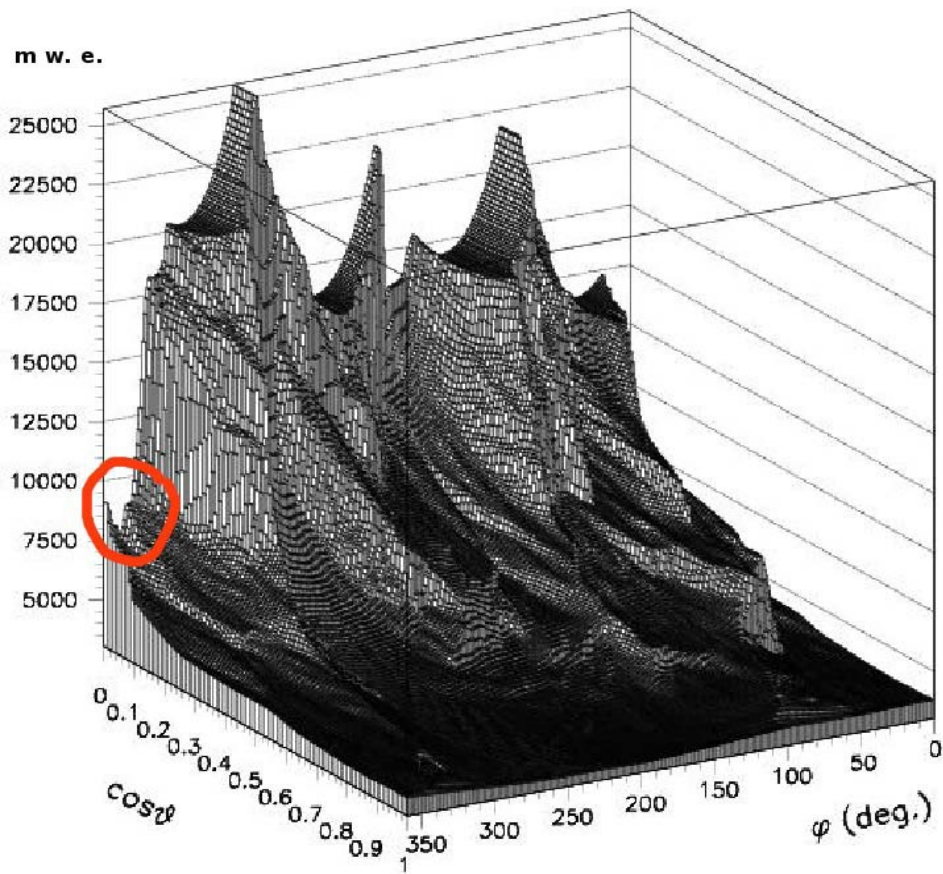

*Fig. 12. Map of the slant depth of the Gran Sasso mountain (in m. w. e.) as a function of the arrival direction: θ and φ are respectively the zenith and azimuth angle. The red circle indicates the direction of the "Teramo valley", where the mountain profile exhibits an "anomaly" in the m. w. e. for horizontal directions.*

Visual inspection using the event displays of both the experiments confirms the hypothesis. Only one event was discarded as due to an un-tagged beam event. Seven events were conservatively discarded as containing too few hits in OPERA. At the end 306 events were used for this analysis (242 from TT, 64 from RPC OPERA sub-systems). As an example we show in Figures 13a and 13b one of such events as seen by the OPERA and LVD event displays. On average 0.27 "Teramo muons" per day are detected. These high-energy muons are well suited to cross-calibrate LVD and OPERA timing systems through a time of flight (TOF) technique totally independent from the (TOF) CNGS neutrino beam.

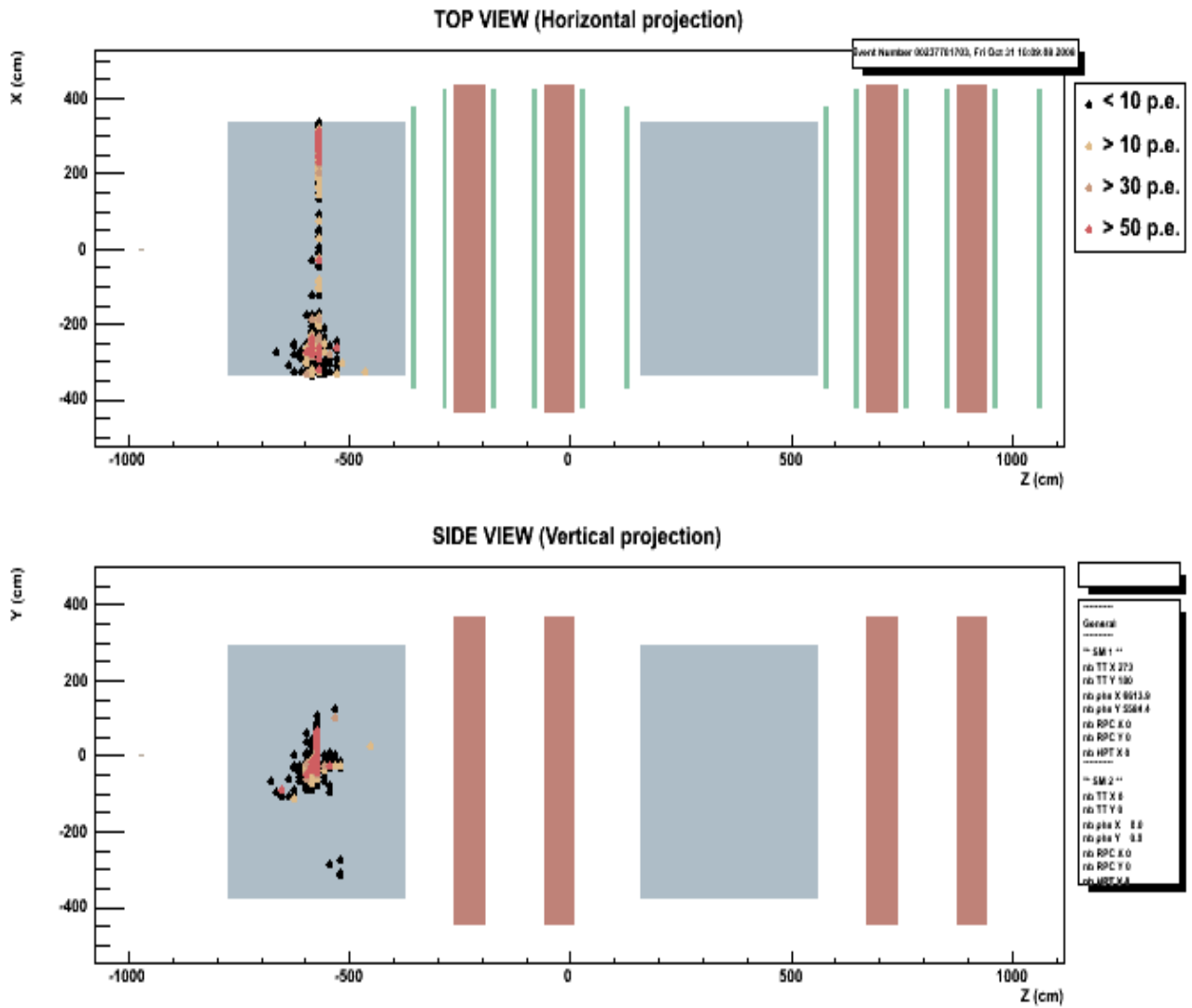

*Fig. 13a. Event display of one of the 306 "Teramo muon" events used for this analysis as "seen" by OPERA.*

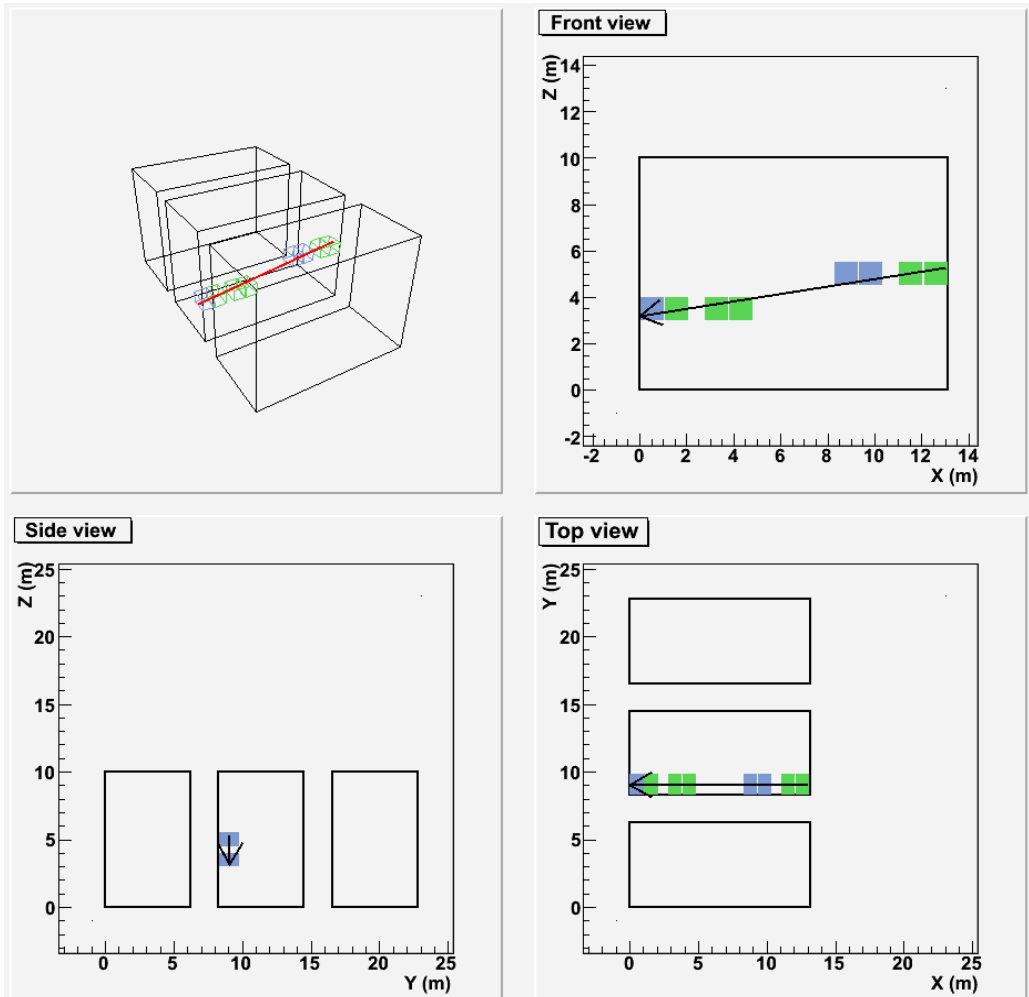

*Fig. 13b. Event display of the same "Teramo muon" event shown in Figure 13a, as "seen" by LVD.*

### *Baseline evaluation*

A preliminary evaluation of the distance between the two detectors was carried out. In the present work the baseline is only used to roughly estimate the expected muon TOF. In the next future it will be used to precisely compare the expected with the measured TOF values.

The perpendicular distance between the two "eastern" sides of both detectors was evaluated using a map of LNGS and it was 160 m (with an uncertainty of about 2 m). The assumption that the digits of the earliest hits are located in the most "eastern" sides of both detectors was checked and corrected for. The average distance of earliest hit from the eastern side of OPERA is 0.4 m with an RMS of 1.0 m. Finally we also corrected for the angle of each muon, which systematically increases the baseline. We computed the exact directions for a subsample of muons, and obtained an additional offset of 2.1 m. The final result gives for the baseline the value of (162±2) m.

## 6. Evidence for a time shift in the OPERA setup

Recently the OPERA Collaboration found a mismatch between the frequency of the internal Master Clock oscillator with respect to the nominal one [8]. We therefore expect a drift of the OPERA time-stamp within the DAQ cycle, which lasts 0.6 s. Since cosmic ray muons are uniformly distributed within the DAQ cycle a dependence of the "Teramo muon" time-of-flight is also expected. This was found to be the case as shown in Figure 14. The fitted value (114±14 ns/s) is compatible with what measured with direct methods (124.08 ± 0.08 ns/s). Each event was therefore corrected according to the formula given below, assuming that the frequency offset measured in 2012 was stable since 2007:

$$t^*_{OPERA} = t_{OPERA} - 124.08 \times 10^{-9} \, t_{DAQ} \qquad (1)$$

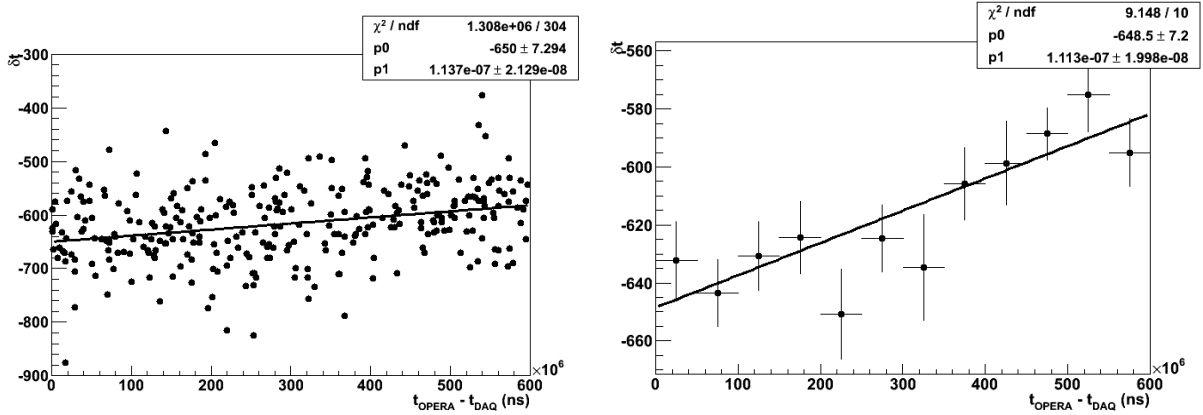

*Fig. 14. Dependence of the "Teramo muons" TOF on the time of the events within the OPERA DAQ cycle. The same events are shown as a scatter plot on the left and a profile histogram on the right.*

We now define the corrected time difference $\delta t = t_{LVD} - t^*_{OPERA}$. The scatter plot of $\delta t$ as a function of the calendar time is finally reported in Figure 15.

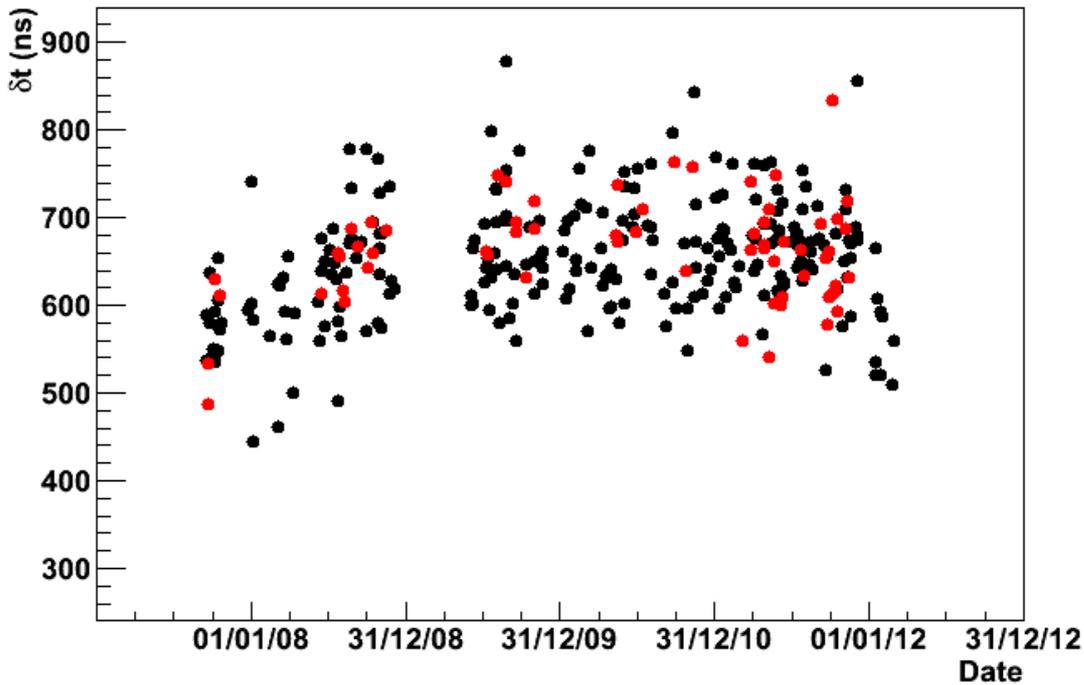

*Fig. 15. Distribution of the $\delta t = t_{LVD} - t^*_{OPERA}$ for corrected events. The black (red) dots represent events originated in the TT (RPC) OPERA sub-system.*

In order to study the stability of the time difference δt versus calendar time, the data have been subdivided in different periods of the various solar years. The year 2008 has been divided in three samples: before May, May-August, after August. For each period we look at the δt distribution, compute the mean and the RMS. The results are shown in Figures 16 and 17 and summarized in Table 2. The total number of events, 306, is distributed into eight samples, each one covering a given calendar time period.

The data show that the average value of δt in each period, goes from ~ 580 ns in 2007 up to ~ 670 ns in the period from August 2008 to 2009, 2010 and 2011; then in the 9 events for the first months of 2012 it decreases again to ~ 570 ns. The observed variations are larger than the statistical uncertainty estimated for each period.

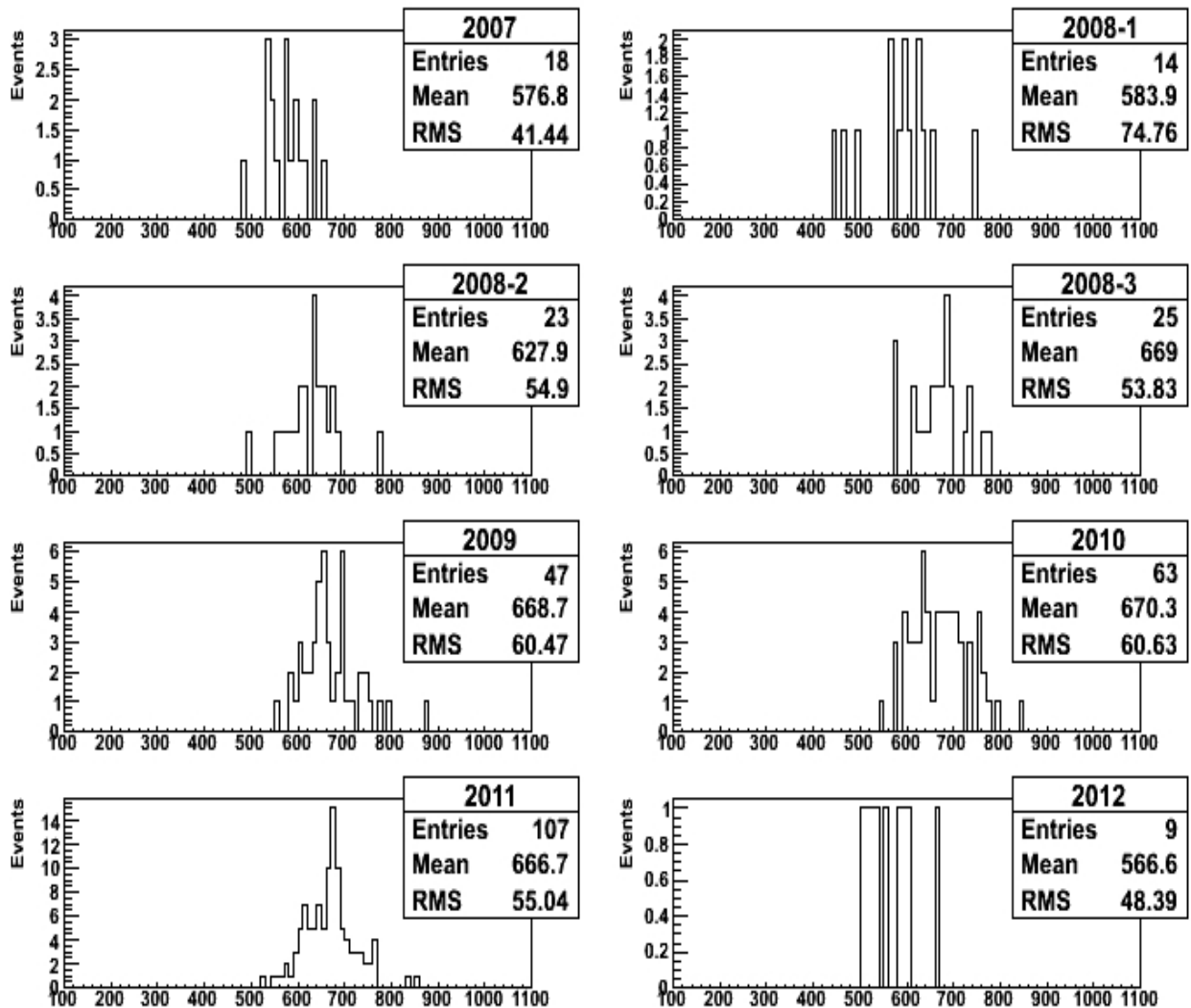

Fig. 16. Distribution of the $\delta t = t_{LVD} - t^*_{OPERA}$ for each period of time.

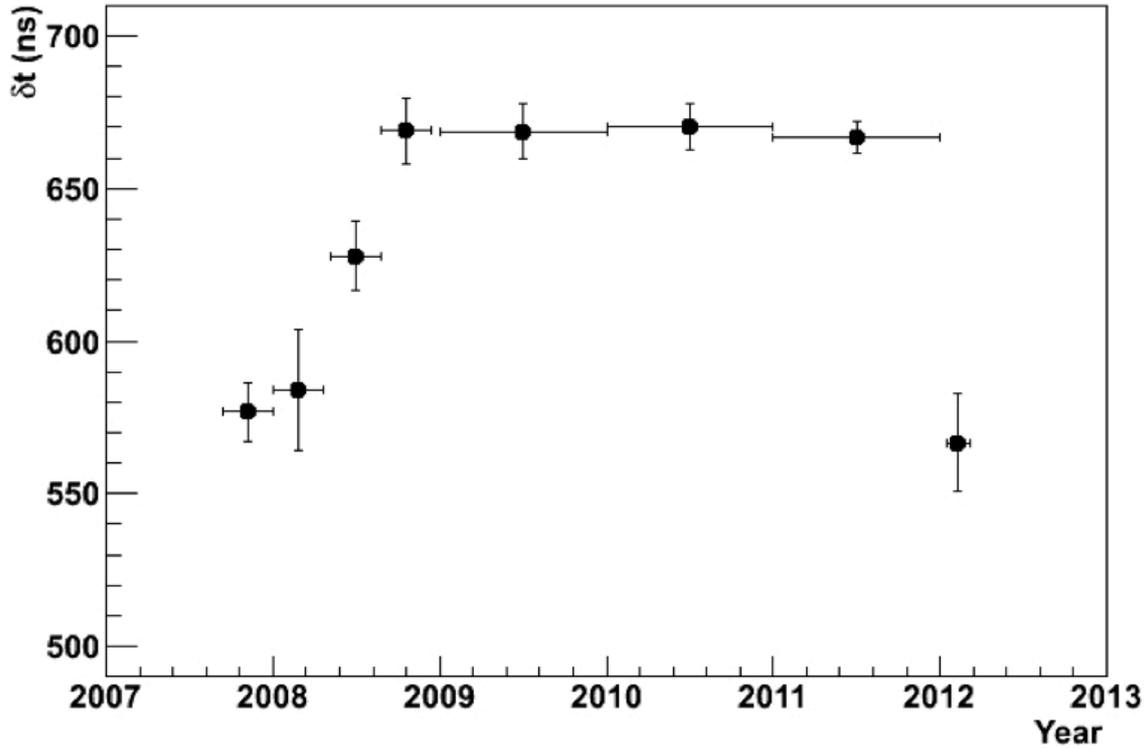

*Fig. 17. Distribution of the δt = $t_{LVD} - t^*_{OPERA}$ for corrected events. All the events of each year are grouped in one single point with the exception of year 2008 which is subdivided in three periods: before May, May-August, after August.*

| TOTAL NUMBER OF EVENTS = 306 | | | | | |
|---|---|---|---|---|---|
| Class | Year | Since | To | Nb. of events | <δt> (ns) |
| A | 2007 | Aug | Dec | 18 | 577± 10 |
| A | 2008-1 | Jan | Apr | 14 | 584 ± 20 |
| A | 2008-2 | May | Aug | 23 | 628 ± 11 |
| B | 2008-3 | Sep | Dec | 25 | 669 ± 11 |
| B | 2009 | Jun | Nov | 47 | 669 ± 9 |
| B | 2010 | Jan | Dec | 63 | 670 ± 8 |
| B | 2011 | Jan | Dec | 107 | 667 ± 5 |
| A | 2012 | Jan | Mar | 9 | 567 ± 16 |

*Table 2: Summary of the δt distribution in the various calendar time periods.*

Let us now group the results in two classes:

- class A: between August 2007 to August 2008 and from January to March 2012;
- class B: from August 2008 to December 2011.

The two separated projected distributions for class A and B are reported in Figure 18. We obtain for class A δt = (595 ± 8) ns, while for class B δt = (668 ± 4) ns. In Figure 19 we report the average value <δt> for the two classes. The resulting time difference between the average values in the two classes is $\Delta_{AB} = <\delta t_A> - <\delta t_B> = (-73 \pm 9)$ ns, far from zero at 8-sigma level. We also note that now, after doing all the needed corrections, the two Gaussian distributions have a width compatible with the ~50 ns time accuracy claimed by the experiments.

The stability in time of LVD shows that the OPERA detector has a negative time shift in the calendar period from August 2008 to December 2011 of the order of $\Delta_{AB} = (-73 \pm 9)$ ns compared with the calendar time from August 2007 to August 2008 and from January to March 2012 taken together.

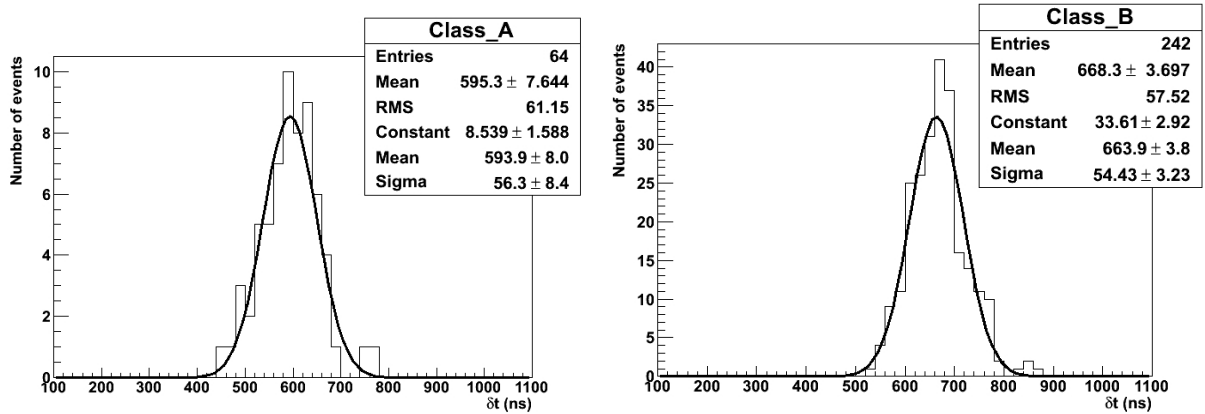

Fig. 18. Distribution of $\delta t = t_{LVD} - t^*_{OPERA}$ for events of class A (left) and B (right).

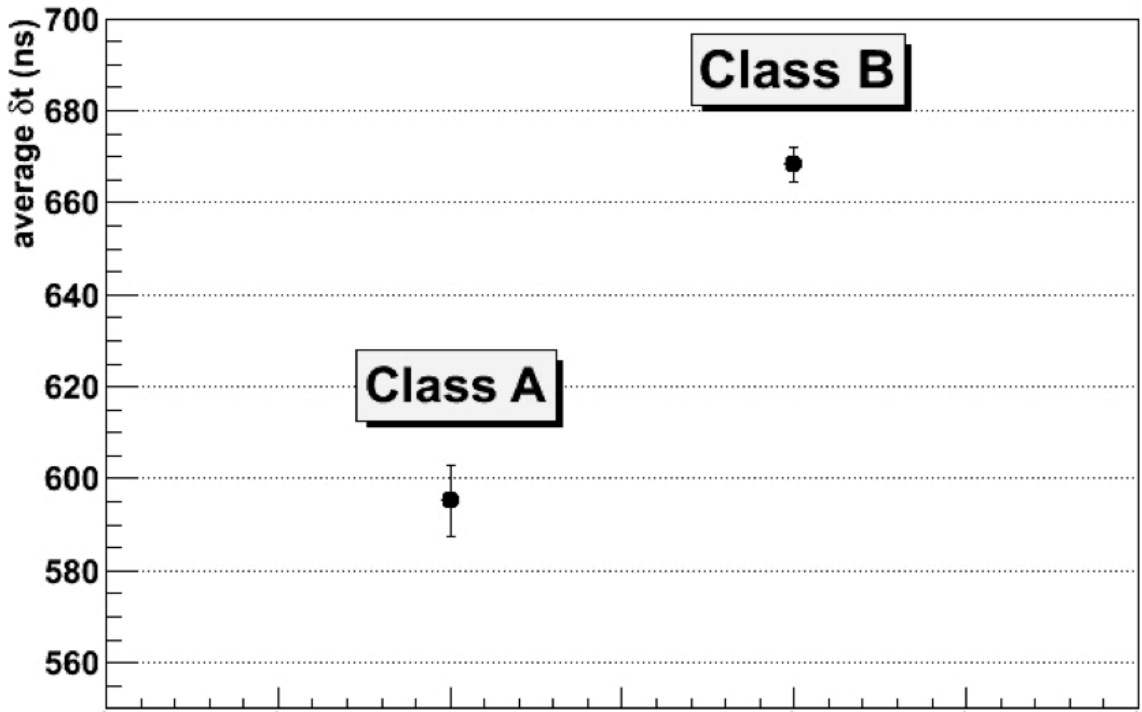

Fig. 19. Average value of $\delta t$ computed in each class of events. Class A are events in calendar time from August 2007 to August 2008 and from January 2012 to March 2012; Class B are from August 2008 to December 2011.

## 7. Summary and conclusions

Data from horizontal muons traversing the LVD and OPERA detectors cover a calendar time period from mid 2007 until 2012, for a total live time of about 1200 days.

In a time-window of 1 μs, and excluding events in time with the CNGS beam spill, we found 306 events due to horizontal muons from the "Teramo anomaly".

This sample has a time-difference ($t_{LVD} - t_{OPERA}$) distribution peaked at 616 ns with an RMS of 74 ns. The central value of the distribution has the following interpretation: the coincident events detected up to now are not multiple muons (one per each detector), but single muon events entering horizontally from the OPERA side and going through the LVD detector with a delay of 616 ns (the delay is a sum of Time of Flight plus a stable systematic error). Indeed, the OPERA-LVD direction lies along the so-called "Teramo anomaly", where the mountain profile exhibits an anomaly in the m. w. e. depth in the horizontal direction. Visual inspection using the event displays of LVD and OPERA detectors confirms this anomaly discovered by LVD in 1997 [1].

The calendar time evolution of the time difference δt for various periods of data-taking is shown in Figure 17. We see an evolution of the average value in each period, ranging from ~ 580 ns in 2007 up to ~ 670 ns from May 2008 to the end of 2011; then for the 9 events collected so far in 2012 it decreases again to ~ 570 ns. The observed variations are larger than the statistical uncertainty estimated for each period.

Grouping the time periods in two classes, as labelled in Table 2, we obtain for class A an average value of Δt (A) = (595 ± 8) ns, and for class B Δt (B) = (668 ± 4) ns. The time stability of LVD compared with that of the OPERA detector gives a time difference between the two classes Δt (A − B) = (−73 ± 9) ns. This corresponds to a negative time shift for OPERA in the calendar period from August 2008 to December 2011 of the same order of the excess leading to a neutrino velocity higher than the speed of light as reported by OPERA [7].

Recent checks of the OPERA experimental apparatus showed evidence for equipment malfunctionings [8]. A first one is related to the oscillator used to produce the event time-stamps, while the second one is linked to the connection of the optical fiber bringing the GPS signal to the OPERA master clock. This allows to conclude that the quantitative effect of this malfunctioning is the negative time shift, Δt (A − B), mentioned above. This explains the previous OPERA finding [7] on the neutrino time of flight shorter by 60 ns over the speed of light.

The result of this joint analysis is the first quantitative measurement of the relative time stability between the two detectors and provides a check that is totally independent from the TOF measurements of CNGS neutrino events and from the effect presented in [8], pointing to the existence of a possible systematic effect in the OPERA neutrino velocity analysis.

If new experiments will be needed for the study of neutrino velocities they must be able to detect effects an order of magnitude smaller than the value of the OPERA systematic effect.

## Acknowledgements

For the measurements reported here we are indebted to LNGS-INFN services, in particular the Computing and the Electronics Services. The Electronics Services of IPHC-Strasbourg, IPNL-Lyon and Bologna-INFN are also warmly acknowledged.